%% file: RR-BJPS-2026-prefinal.tex
\documentclass[11pt,a4paper]{article}
\usepackage{hyperref}
\usepackage{tikz-cd}
\usepackage[utf8]{inputenc}
\usepackage{graphicx}  \usepackage{mathptmx}
\usepackage{amsmath, amssymb}
\numberwithin{equation}{section}
\usepackage{gensymb}

\input{Bohrificationpreamble.tex}
\newtheorem{definition}{Definition}[section]

\newtheorem{theorem}[definition]{Theorem}

\newtheorem{corollary}[definition]{Corollary}
\newcommand{\ML}{Martin-L\"{o}f}

\topmargin = - 1 cm \textheight = 23 cm \textwidth = 15 cm
\usepackage[symbol]{footmisc}
\renewcommand{\thefootnote}{\fnsymbol{footnote}}

\newcommand{\QED}{\null\nobreak\hfill\ensuremath{\square}}%

\begin{document}
\pagenumbering{arabic} \setlength{\unitlength}{1cm}\cleardoublepage
\date\nodate
\begin{center}
\begin{Huge}
{\bf Randomness from  initial conditions: \vspace{10pt} \\  The challenge of incompleteness}\footnote{This paper was inspired by the \emph{Workshop on Algorithmic Randomness and Quantum Mechanics} in Cracow, August 31 and September 1, 2026, and takes a different direction from my talk given there. I am grateful to Joanna Luc and Tomasz Placek for organizing this  meeting, and thank many of its participants for questions and discussions. I am also indebted to Jeffrey Barrett, Eddie Chen, and Nino Dekkers for comments on an earlier draft of this paper, and to Mathieu Hoyrup for correspondence. Finally, Claude Opus 5.5  ``refereed'' an earlier draft and found several errors.}
\end{Huge}
\bigskip\bigskip

\begin{Large}
 Klaas Landsman\footnote{ Radboud Center for Natural Philosophy (RCNP) and Department of Mathematics, Institute for Mathematics, Astrophysics, and Particle Physics, 
Radboud University,  Nijmegen, The Netherlands. \texttt{landsman@math.ru.nl}}
 \end{Large}\medskip
 \medskip

 \begin{abstract} 
\noindent
Certain \emph{prima facie} deterministic theories, notably 
Bohmian mechanics and kinetic theory, require ``random'' initial conditions in  order to achieve their goals. I analyze recent proposals that this  idea should be implemented using algorithmic (\ML) randomness,  seen as a mathematical sharpening of 
 the  typicality arguments originally used to understand these theories. On the one hand my analysis includes a critique of these arguments and hence a defense of this sharpening. On the other hand, I argue that Chaitin's incompleteness theorems for random sequences make it impossible for a dynamical theory to either process \emph{given} random initial conditions beyond some finite (but \emph{not} arbitrary) number of their digits, or else to \emph{state} such conditions with arbitrary precision without compromising their randomness (these are two sides of the same coin).
 This  challenges the ability of such theories to make any other predictions than those that hold typically in the usual sense of the word (like macroscopic statements),
  and moves their alleged determinism high up the metaphysical ladder.  \end{abstract}\end{center}
 \tableofcontents
 
\thispagestyle{empty}
\renewcommand{\thefootnote}{\arabic{footnote}}
\newpage \setcounter{footnote}{0}
\section{Introduction}
An interesting recent development in the philosophical and mathematical study of (in)determinism has been the proposal
 to impose algorithmically random initial conditions in  theories whose dynamics is deterministic.  This has specifically been done, for different purposes, 
 in both Bohmian mechanics (Barrett,  Chen, and  Lopez-Wild, 2026ab) and kinetic theory (Dekkers and Landsman, 2026),
  but this move surely has broader implications for both classical and quantum physics.   
 
 In the first case, the goal of the proposal was to give a sound mathematical and philosophical underpinning of a (if not \emph{the}) fundamental claim of Bohmian mechanics,\footnote{For this claim and its context see also D\"{u}rr and Teufel (2009), Barrett (2019), and Goldstein (2025).}  namely that:
\begin{quote}
\begin{small}
The origin of the randomness in the results of quantum measurements lies in random initial
conditions, in our ignorance of the complete description of the system
of interest---including the apparatus---of which we know only the wave
function. (D\"{u}rr,  Goldstein, \& Zangh\`{\i}, 1992, p.\ 844)
\end{small} 
\end{quote}
This proposal provides an exact mathematical definition of the intuitive idea that the initial particle positions $(q_0, \ldots, q_{N-1})$  `are ``randomly distributed'' according to the quantum equilibrium distribution' (\emph{ibid.}, p.\ 856), i.e., the Born probability measure $P=|\Psi|^2$ on $\R^{3N}$ (at the initial time),\footnote{Here the informal notation $P=|\Psi|^2$ is shorthand for $dP(q_0, \ldots, q_{N-1})=|\Psi(q_0, \ldots, q_{N-1})|^2dq_0\cdots dq_{N-1}$.}
 where $\Psi(q_0, \ldots, q_{N-1})$ is the wave-function of the  universe. Their definition states that the initial positions are $|\Psi|^2$-random (in the  sense of \ML), from which Barrett,  Chen, and  Lopez-Wild (2026ab) showed  that outcome sequences of infinite runs of measurements as described by Bohmian mechanics are also random relative to a single-particle  Born measure $|\psi|^2$, which in turn implies that outcome probabilities defined as relative frequencies satisfy the usual Born rule.

In kinetic theory,\footnote{Sklar (1993) and Uffink (2007) remain  excellent conceptual introductions;
for Boltzmann see also  Uffink and Valente (2015), Darrigol (2018), and Uffink (2022). 
 The  mathematical frontier is Deng, Hani and Ma (2024). Prior to Dekkers and Landsman (2026),  Hiura and  Sasa (2019)  pioneered the use of algorithmic randomness in kinetic theory. } the motivation for proposing algorithmically random initial conditions was to make precise sense of Boltzmann's assumption of ``molecular chaos'' \emph{for individual gases} (rather than ensembles). In the specific form that two particles are uncorrelated before they collide, this 
assumption (which also had roots in Maxwell's work) was essential for Boltzmann's derivation in 1872 of the transport equation named after him,
 and it has remained a central ingredient in all subsequent work on the Boltzmann equation. This time, the background measure $P$ is defined on the set $X=\R^{6N}$ of all positions \emph{and momenta}  $x=(p_0, q_0, \ldots, p_{N-1}, q_{N-1})$, and  is  given  by\footnote{For simplicity we omit terms taking care of the fact that hard spheres cannot overlap.}
\begin{equation}
dP(p_0, q_0, \ldots, p_{N-1}, q_{N-1})=\prod_{n=0}^{N-1} f_0(p_n,q_n) dp_ndq_n. \label{iid}
\end{equation}
Here $f_0(p,q)$, which is the same function for all $n$,  is a probability density on the single-particle phase space $\R^6=\{p,q\}$,
so that  $P$ makes all particles independent and identically distributed (i.i.d.) at $t=0$. The goal, which Dekkers and Landsman (2026)  achieved in two toy models, is then to show that $P$-random  \emph{microscopic} configurations $x$ at $t=0$ satisfy Boltzmann's assumption, so that in the (Boltzmann--Grad) limit $N\raw\infty$
the \emph{macroscopic} single-particle distribution function
\begin{equation}
f_t (p,q; p_0, q_0, \ldots, p_{N-1}, q_{N-1})= \frac{1}{N}\sum_{n=0}^{N-1} \dl(p-p_n(t))\dl(q-q_n(t))\label{1pd}
\end{equation}
only depends on $(p,q)$ and then solves the Boltzmann equation with initial value $f_0$.
 
 The main idea of algorithmic randomness (as I see it) in the present context is as follows. 
  Suppose  one has a probability space $(X,\Sg,P)$, where $X$ is a set, $\Sg$ a collection of measurable subsets of $X$ (technically a $\sg$-algebra), and $P:\Sg\raw[0,1]$ a probability measure on $X$, all assumed computable (or `effective') in a suitable sense.\footnote{See Hertling and  Weihrauch (2003) or  Hoyrup and  Rojas (2009) for definitions and details.}
Here $X$ is typically (\emph{sic}) a space of \emph{microscopic} degrees of freedom,
such as $X=\R^{3N}$ in Bohmian mechanics, $X=\R^{6N}$ in kinetic theory, or $X=2^N$ seen as a toy model for either,\footnote{Here $N=\{0,1, \ldots, N-1\}$ \`{a} la von Neumann, so that  $2=\{0,1\}$, and  $2^N$ is the set of binary strings of length $N$. One may  replace $N$ by its limit set $\N=\{0,1,2,3, \ldots\}$, so that  $2^\N$ (often called $2^{\omega}$) is the set of infinite binary sequences.
 } or simply as the set of all possible outcomes of an $N$-fold coin toss. We will often idealize $N$, supposedly huge, by $\N$.
We already specified $P$ for our two case studies; for an infinite coin toss we take the Bernoulli measure $P=f^\N$ on $2^\N$, defined on cylinder sets by 
\begin{align} 
f^\N(\sg 2^\N)=2^{-N} && (\sg\in 2^N),
\end{align} 
where $\sg 2^\N$ is the set of all infinite binary sequences that start with the given binary string $\sg$ of length $N$ (note that this Bernoulli measure  assigns zero probability to any sequence $s\in 2^\N$).\footnote{The cylinder sets $\sg 2^\N$, where $\sg\in 2^N$, generate the $\sg$-algebra $\Sg$ on which $f^\N$ is defined. The notation $f^\N$ comes from the uniform probability $f$ on $\{0,1\}$, given by $f(0)=f(1)=1/2$, which is infinitely repeated to yield $f^\N$.} 

Furthermore, typicality reasoning in whatever form usually concerns some \emph{macroscopic} property
 $M$, also written as $M(x)$, which logically speaking is a predicate on $X$; that is, a well-defined property that elements $x\in X$ may or may not have, i.e., $M(x)$ may be true or false depending on $x$. For example, for the infinite coin toss with $X=2^\N$ the predicate  $M(x)$ could be the property
 \begin{align}
 \lim_{N\raw\infty}\frac{1}{N}\sum_{n=0}^{N-1} x_n=1/2 && (x=x_0x_1x_2\cdots,\, x_n\in \{0,1\}). \label{SLLN}
 \end{align}
In Bohmian mechanics $M(x)$ would state that  initial particle positions $x=(q_0, \ldots, q_{N-1})\in \R^{3N}$ induce random outcomes of measurements that satisfy the Born rule, and in kinetic theory it would be the property that 
a gas with initial state $x= (p_0, q_0, \ldots, p_{N-1}, q_{N-1})$ in phase space $X=\R^{6N}$ 
  satisfies the Boltzmann equation with some initial condition  $f_0$ (and hence evolves irreversibly). 
  
As discussed in some more detail in \S\ref{S2}, a common pattern of explanation in the philosophy of physics is that
$M$ occurs because it holds ``typically'', that is,  $M(x)$ is true for ``an overwhelming number of  $x\in X$''. In my view,\footnote{See also Barrett and Chen (2025), Barrett,  Chen, and  Lopez-Wild (2026ab), and Wilhelm (2025a).}
this is too vague for mathematical physics, but in most cases
 such arguments can be made precise by choosing some prior probability measure $P$ on $X$. The proposition that
 $M$ holds typically then means that 
$M(x)$ is true for $P$-almost every $x\in X$ (i.e., with probability 1); in other words, 
there is some subset $\mathcal{N}\subset X$ with $P(\CN)=0$ such that  $M(x)$  for all $x\in X\backslash\CN$.
There are also many such situations in mathematics;
the strong law of large numbers, which states that \er{SLLN} holds typically with respect to $P=f^\N$, is a good example, as is Birkhoff's pointwise ergodic theorem.
But no \emph{explicit} atypical set $\CN$ is  described in  ``typicality theorems''! In other words, typicality arguments provide no information about \emph{which} $x\in X$ satisfy $M(x)$, and hence give at best incomplete understanding of the occurrence of the property $M$ in  \emph{this} world.\footnote{Many philosophers indeed claim that this requires no explanation once $M$ is typical, cf.\ \S\ref{S2} and references therein.}

Moreover, in our case studies and elsewhere, one would like to say not only that a (macroscopic) \emph{property} $M$ holds typically with respect to $P$, 
but also that some (microscopic) \emph{point} is typical with respect to $P$. For example, for a coin toss  the average outcome being (approximately) 1/2 is a far coarser indication that the outcome sequence is not rigged than this sequence
being ``random'' or ``chaotic''. In
kinetic theory one  would like the  initial configuration of a gas to be  random with respect to the i.i.d.\ measure \er{iid}, and 
Bohmians often say that the initial particle configuration in the universe is random with respect to the Born measure $P=|\Psi|^2$, for example:
\begin{quote}
\begin{small}
To create a universe God must first decide upon (\ldots) 
a particular realization, out of all possible solutions, of the dynamics--the
one corresponding to the actual universe. (\ldots)  the simplest
possibility would appear to be a completely random initial condition, with
an appropriate natural measure for the description of this randomness
(whatever this might mean, even given the measure). (D\"{u}rr,  Goldstein, \& Zangh\`{\i}, 1992, p.\ 896)
\end{small} 
\end{quote}
In other words, one tries to eat one's cake and have it, in attempting to
resolve  the tension between on the one hand wishing to state that some microscopic configuration $x\in X$ is typical \emph{with respect to some macroscopic property $M$}, whilst on the other hand saying that $x$ is typical \emph{with respect to some probability measure $P$ on $X$}. 
The notion of typicality sought for  is precisely what is expressed by the concept of $P$-randomness in the sense of \ML\ (which is a particular flavour of algorithmic randomness that suffices for my purposes).
See \S\ref{S3} for details.  Indeed, $P$-randomness is by definition a property that $x\in X$ may or may not have, and in addition
 it  may resolve the first issue of explicating the typical set, since one may often 
 sharpen some theorem of the generic form 
 \begin{center}
 \emph{$M(x)$ holds for  $P$-almost every $x\in X$,}
 \end{center}
 that is, $M$ holds typically with respect to $P$, 
 by a corresponding theorem stating that
  \begin{center}
 \emph{$M(x)$ holds for  every $P$-random $x\in X$}.
 \end{center}
 Since under very mild conditions the set $\mathrm{Ran}_P(X)$ of $P$-random points in $X$ has probability one,\footnote{This is the case whenever
 $(X,\Sg, P)$ is an effective topological measure space,  which is true  
 in all interesting applications. See Hertling and Weihrauch (2003), Definition 3.1 and Proposition 3.11.1.}
 \begin{equation}
P(x\mbox{ is $P$-random})\equiv P(\mathrm{Ran}_P(X))=1, \label{P1}
\end{equation}
this indeed sharpens the original theorem in providing an explicit ``typical set'' on which $M$ holds, namely $\mathrm{Ran}_P(X)$. As said, this sharpening often works: the simplest and oldest example is  Borel's strong law of large numbers, refined by stating that every $f^\N$-random  $x\in 2^\N$ satisfies \er{SLLN}.\footnote{See \ML\ (1966), as well as Dekkers and Landsman (2026), \S 1, for a pedagogical treatment.}
 More generally, effective versions of Birkhoff's pointwise ergodic theorem along these lines have also been derived,\footnote{See Towsner (2020) for a review and references. In the traditional form this theorem implies Borel normality of almost every real in $[0,1]$ with respect to Lebesgue measure $\lambda$, whereas its algorithmically random form yields this property for every $\lambda$-random real in $[0,1]$. This, in turn, implies the amazing ``monkey typewriter theorem'': the decimal expansion of each $\lm$-random real in $[0,1]$ contains  every finite number infinitely often, cf.\ Calude (2002), Theorem 6.50. }
and in two well-known toy models the usual ``almost everywhere'' derivation of the Boltzmann equation has also been refined in the said way.\footnote{See Dekkers and Landsman (2026).}  Finally, Barrett,  Chen, and  Lopez-Wild (2026ab) accomplished this sharpening for the Born rule in Bohmian mechanics.\footnote{The original version of the theorem was due to D\"{u}rr,  Goldstein, and Zangh\`{\i} (1992), at least heuristically.}

Note that $P$-randomness, like our earlier $M$, is  a predicate $\mathrm{Ran}_P$ on $X$, defined by stipulating that 
$\mathrm{Ran}_P(x)$ is true iff $x$ is $P$-random, so that $\mathrm{Ran}_P(X)=\{x\in X\mid \mathrm{Ran}_P(x)\}$.
One difference with $M$ is that in most applications $M$  expresses some macroscopic property that somehow averages out over $x$, whereas $\mathrm{Ran}_P$ concerns $x$ in its microscopic details. Another difference is that unlike our $M$'s so far, $\mathrm{Ran}_P$ depends on the probability measure $P$ against which typicality is judged!

However, interpreting typicality with respect to some probability $P$ as $P$-randomness faces a serious challenge in the form of \emph{Chaitin's incompleteness theorems}, which will be 
 reviewed in \S\ref{S3}. Roughly speaking, these theorems imply that 
random points (which for us are initial conditions) cannot be stated in sufficiently precise numerical form.
This  harbors new problems for both Bohmian mechanics and kinetic theory, in which respect the former seems more vulnerable. Although in  kinetic theory  undecidability  elevates its original inability to compute microscopic trajectories \emph{in practice}  to an inability to do so \emph{in principle}, this seems acceptable since it never had that goal in the first place. But in Bohmian mechanics, the particle positions are all it adds to QM (plus a new role of the wavefunction in steering these positions as a ``pilot wave''), and so their inaccessibility not just in practice (after all they are hidden variables) but even in theory may be less acceptable: the  positions `\emph{can be known of, but not known through human reason.}'\footnote{Paraphrasing Bennett and Gardner's (1979) famous words about Chaitin's random number $\Om$.}

Here is the plan of the remainder of the paper.  In \S\ref{S2} I critically discuss typicality as currently viewed by an influential community in  the philosophy of physics,\footnote{See for example Bricmont (2016, 2022), Maudlin (2020), Lazarovici (2020, 2023),
and Wilhelm (2025abc, 2026). Despite some differences between these authors, one may distill what I see as a ``received view'' from this literature, based on the idea that typicality provides a way of arguing and explaining that is different from probabilistic reasoning and even independent of it in some cases, and is also  successful both in kinetic theory and in Bohmian mechanics.
 \label{Laza}}
 and thereby motivate  its possible strengthening by algorithmic randomness--not just, but especially in Bohmian mechanics and Boltzmannian kinetic theory. In \S\ref{S3} I review the  incompleteness theorems, which in 
 \S\ref{S4} is applied to the case of random initial conditions. Incompleteness makes these conditions and their ensuing trajectories even more inaccessible than they already were, which I argue to have impact on the concept of determinism and the question whether  Bohmian mechanics is really deterministic.
\section{Rethinking typicality}\label{S2}
Let us return to the setting where $X$ is a set, which  may be thought of as 
the set $X=2^\N$ of infinite binary sequences, seen as the outcomes of infinite coin tosses, or as toy models for  $X=\R^{3N}$ (Bohmian mechanics) or $X=\R^{6N}$ (kinetic theory), both seen as initial $N$-particle configurations at say $t=0$, about to deterministically evolve  according to Bohmian or  Newtonian dynamics, respectively (both with respect to some given potential), possibly for $N\raw\infty$. 
I only consider cases where points $x\in X$ are seen as \emph{microscopic} states.
Furthermore, let $M$ be a predicate on this set, as discussed in the Introduction,  that is, a property (if you like, expressed by a formula $M(x)$ in first-order logic interpreted in $X$) that points $x\in X$ may or may not have, so that $M(x)$ may or may not be true for given $x\in X$. Here one usually thinks of $M$ as a \emph{macroscopic} property.

The idea of typicality, then,  is that our property  $M$ holds in the world because it is ``typical'':\begin{quote}
\begin{small}
\textbf{Definition.} Let $X$ be a reference set (\ldots)
A property $M$ is typical within $X$ if nearly all members of $X$ instantiate $M$. (Lazarovici, 2020, p.\ 2)\footnote{
Notation adapted to mine.}

\end{small} 
\end{quote}
So far so good, especially in clarifying that primarily it is \emph{properties}, i.e., predicates on $X$,  that may or may not be typical, rather than \emph{points} of $X$. But what does `nearly all members of $X$' mean? It is safe to say that the practice of modern mathematical physics is univocal in interpreting this in the following way: typicality must be defined with respect to some probability measure on $X$ (more precisely, on some $\sg$-algebra $\Sg$ on $X$), in which case $M$ is typical if it holds $P$-almost everywhere with respect to $P$; in other words, if there is a null set $\CN\subset X$ (i.e., $\CN\in\Sg$ and $P(\CN)=0$)
such that $M(x)$ is true for all $x\notin\CN$. In my pedagogical example \er{SLLN}  the strong (Borel) law of large numbers says exactly that. This concept of typicality may be weakened (as is practically always needed in finite systems, or, in my example, in $X=2^N$ instead of $2^\N$) by `epsilonizing' this argument, i.e., by taking some $0<\varep<1$  and declaring that $M$ holds typically provided $P(\CN)<\varep$ instead of $P(\CN)=0$.\footnote{For an interesting example see Theorem 1 in Chen and Tumulka (2026).} But this does not affect the point that at least in mathematical physics: 
\begin{center}
\emph{probability measures come before typicality}.
\end{center} For example, the proof of the Boltzmann equation for arbitrarily long times (for which solutions exist) by Deng, Hani, and Ma (2024) is based on this notion of typicality, here with respect to the probability measure \er{iid}, as was most of the preceding rigorous mathematical work on this topic. 
Similarly, as is clear from the definition of \ML\ randomness reviewed in \S\ref{S3}, 
perhaps against Kolmogorov's original intentions,\footnote{Predating the work of his guest \ML, Kolmogorov was one of the founders of algorithmic randomness in the context of computer science, where he studied program complexity. See  Li and  Vit\'{a}nyi (2008) and Porter (2014).} one cannot avoid the mathematical fact that:
\begin{center}
\emph{probability measures come before randomness}.
\end{center}
Ironically, our current (\ML) theory of algorithmic randomness, in which this claim holds, 
 is the outcome of a bumpy process that started with work by von Mises in 1919 that had the opposite goal of basing probability on randomness!\footnote{See van Lambalgen (1987),  Li and  Vit\'{a}nyi (2008), and Porter (2012) for this history.} 
 Thus the undeniable foundational problems with the concept of probability remain, inherited by  typicality and (algorithmic) randomness as discussed so far; what's more, the former is in my view incomplete and should be strengthened by the latter, in which case the challenge addressed in the next section must be added to our headaches. 

Hence one cannot but have sympathy for the more radical approach to typicality currently favoured in 
the philosophy of physics literature,\footnote{At  least as represented by the references in footnote \ref{Laza}.}
 which is based on the opposite order of things:
\begin{center}
\emph{probability and randomness are grounded in typicality}.
\end{center} 
To believe in this possibility, one should first make sense of the above \textbf{Definition}, which hinges on some subdefinition of `nearly all members of $X$'.  As to what this could mean, we read that:
\begin{quote}
\begin{small}
we
don't need to worry too much. In general, all reasonable measures will agree on what
is typical or atypical.
Tim Maudlin (private communication) provides the following instructive example:
The Sahara desert is nearly all sand. One may ask back: ``Nearly all by what measure?
In terms of area, or volume, or metric tons, \ldots?'' But this means that one didn't really
get the point. The answer is: All of them. Or any of them. By any reasonable
standard, the Sahara is nearly all sand. The typical is so overwhelming in number
that it leaves little room for ambiguity. (Lazarovici, 2020, p.\ 3)
\end{small}
\end{quote}
Admittedly, this is made more precise later by relying on so-called \emph{typicality measures}, but in finite sets this is always provided by the counting measure and in the continuum it is claimed that `the Lebesgue measure is always the appropriate typicality measure', at least in both equilibrium and  a `fundamentally deterministic setting'; other cases are claimed to be  `non-equilibrium macro-conditions rather than non-equilibrium typicality measures' (\emph{ibid.}, p.\ 43).\footnote{Neither the quantum equilibrium measure $|\Psi|^2$ in Bohmian mechanics nor the measure \er{iid} in kinetic theory is a Lebesgue measure, which even apart from (surmountable) normalization problems, would not work in either case.} The 
simplest intuition pump on which this view, which is typically attributed to Boltzmann,  is based, is the fair coin toss, where it is indeed true that the number of outcome sequences whose average is close to $\half$ dwarfs all the others as $N\raw\infty$. However, for a biased coin this does not work at all, and in order to state that an outcome sequence is typical (for the given bias) one \emph{has} to refer to the corresponding biased Bernoulli measure, which as $N\raw\infty$, arbitrarily drifts apart from the counting measure. 

 Moreover, in the idealized case $X=2^\N$, even the fair Bernoulli measure $P$ cannot  be said to be an unambiguous representative of any reasonably quantitative way of specifying the ``size'' of a set: it is well known that all probability one sets on which \er{SLLN} holds are topologically \emph{meagre}.\footnote{A subset of a topological space is \emph{nowhere dense} iff the interior of its closure is empty (in other words, if its closure does not contain any nonempty open set), and \emph{meagre} (or \emph{first category})  if it is a countable union of nowhere dense sets. For example, the rationals are meagre in $\R$ (though dense), and  the Cantor set within $[0,1]$ is meagre (but not dense). Frigg (2009) already discussed the contrast between meagre sets and probability 1 sets in the context of typicality, though not quite in the same way as here. For him, the point was that phase space points can be typical (namely probabilistically) in a different way from Hamiltonians (namely topologically). My goal is to highlight potentially conflicting meanings of  mathematically ambiguous expressions like ``overwhelming in number'' or ``ridiculously small''. See also Frigg and Werndl (2012) for a critical analysis of typicality reasoning.}  This is a property that, as the word suggests, indicates that it is ``small'' as a part of $X$; moreover, its atypical complement, which has $P=0$, is topologically \emph{dense!}
 More generally, as we have already seen, under mild assumptions the set  $\mathrm{Ran}_P(X)$ of $P$-random points in $X$ in a computable probability space $(X,\Sg,P)$ has probability one, but once again it is also meagre with dense complement.\footnote{This is the case whenever   an effective topological measure space
 $(X,\Sg, P)$ has a universal \ML-test, which in turn is true if $P$ is merely upper semicomputable, 
cf.\ Hertling and Weihrauch (2003), Proposition 3.11.2.} 

I therefore prefer the following view (still from a committed typicality philosopher):
\begin{quote}
\begin{small}
There are many accounts of typicality, of course: some appeal to cardinality (\ldots); some appeal to purely topological notions (\ldots); and some appeal to other mathematical devices. Each account formalizes a distinct, but still perfectly respectable, notion of what it is for something to be typical. So there is no conflict between the measure-theoretic account and these other accounts, since the notions of typicality that all of these accounts formalize are different. (Wilhelm, 2025a, p.\ 1192) \end{small}
\end{quote}
That said,  I restrict myself in this paper to versions of typicality that are defined by a probability measure; which is all we need for our case studies, and  argue that typicality  can only benefit from the identification of typical points with (\ML) random ones (Landsman, 2023).

 One may also wonder why a property being typical (in whatever sense of the word) is enough to explain its actual occurrence.
 Here   the  ``received view''  has been expressed as follows:
 \begin{quote}
\begin{small}Suppose we accept a theory $T$ and we come to believe that our world
has the salient property $M$ (expressing, for instance, a physical regularity
or phenomenon). If $M$ is typical according to $T$, there is nothing left to
explain.
 (Lazarovici, 2020, p.\ 4)
 
 Explanations have to end somewhere. If we can establish that a
certain property is typical for a particular kind of system, this should remove any sense
of mystery or puzzlement as to why we find such systems instantiating the respective
property. Hence, we should consider the phenomenon to be conclusively \emph{explained} on
the basis of the microscopic theory. (\emph{ibid.}, p.\ 121)
\end{small}
\end{quote}
In other words, no further questions should be asked, much as the Copenhagen interpretation of quantum mechanics--so derided by the typicality philosophers--forbids us to ask what exactly happens during a measurement. But no one can stop me from asking why our macroscopic property $M$ happens in \emph{this world}, and what could be the microstate $x\in X$ of \emph{this world} so that $M$ happens.
This  question about \emph{individual} systems is beyond the range of (philosophical) typicality accounts, which seem to be based on some kind of Lewisian possible world semantics:
 \begin{quote}
\begin{small}
The applications of typicality that we will focus on for the most
part have a decidedly modal character. In particular, the relevant typicality statements
in physics refer, in general, to what obtains in most nomologically possible worlds, not
to what obtains most of the time in the actual universe. 
 (Lazarovici, 2020, p.\ 3)
 \end{small}
\end{quote}
But if I do ask my question, then one might hope that typicality as algorithmic randomness is able to answer it--one explains  $M$ by applying \emph{modus ponens} to  the following conjunction:\begin{center}
1.  $M(x)$ is true for any $P$-random state $x\in X$; \hspace{25pt}
2. The world is in such a state.\end{center}
Here the first part is a matter of mathematics; it says that $x$ instantiates the property $M$ and hence is typical for both $M$ and $P$. The second must be settled by either declaring this to be a law of nature, as proposed for Bohmian mechanics by Barrett and Chen (2025), or by new arguments, as, for now, alas, is the case for kinetic theory. Either way, this \emph{modus ponens}
does not contradict mainstream typicality; it is rather supposed to sharpen both typicality and probability.\footnote{In most cases probability theory is incapable of explaining some individual outcome $x\in X$, since in cases like $X=2^\N$ with Bernoulli measure, and all continuum cases like Bohmian mechanics and kinetic theory, every such outcome has probability zero, and without the idealization, so for $X=2^N$ with $N<\infty$ (etc.), every outcome has the same probability, that is, $x=00000\cdots 00000$ is as likely as $x=10110\cdots 01100$, although the former suggests foul play.} 
\section{The challenge of incompleteness}\label{S3}
\begin{quote}
\begin{small}
Mathematical objects often have a general definition which has no regard for any method or procedure that can describe it.
(Monin, 2020, p.\ 233)
\end{small}
\end{quote}
The overarching problem with random initial conditions, indeed the problem with any stipulation of randomness, is that it is elusive. First, randomness has no universal defining property or irreducible core;  it seems to be a \emph{family resemblance}, with the further complication that any specific family member is defined \emph{negatively}, i.e., through its antipode (Landsman, 2020). For example, in Aristotle a random event is the polar opposite of an event planned towards a specific goal.
If randomness is a form of (human)  ignorance, then its antipode is complete (``God-like'') knowledge.  If it is seen as unpredictability, then its antipode is predictability; if it is seen as indeterminism, then it is antipolar to determinism--although, as we shall see, it is a very subtle question what \emph{that} should mean! If randomness is lack of causation (\emph{pace} Russell), then, as claimed in the Copenhagen interpretation of quantum mechanics,  at least the randomness of measurement outcomes means, besides some probabilistic formalism, that they escape the causal chain of events once thought to rule the world.
 \emph{Et cetera}. These different meanings of randomness sometimes overlap, as in quantum theory being both acausal and unpredictable, and sometimes contradict, as in chaotic dynamical systems being simultaneously deterministic and unpredictable. 
Better than overlapping:  \ML\ randomness may alternatively be defined via each of three \emph{equivalent} and appealing antipodes.\footnote{The equivalence  $1\lraw 2$,  the most spectacular one, is Theorem 6.35 in Calude (2002) or Theorem 3.6.1 in  Li and  Vit\'{a}nyi (2008). For $1\lraw 3$ see Li and  Vit\'{a}nyi (2008), Corollary 4.5.3, or Nies (2009), Proposition 7.2.6. See Downey and Hirschfeldt (2010) for all three.
The seductive argument that \ML\ caught the true nature of randomness because his definition has all these different faces, which the author also finds compelling, is analyzed in Porter (2021). 
} 
 Namely, an infinite binary sequence $x\in 2^\N$ is $P$-random for $P=f^N$ iff it is:\footnote{ Terwijn (2016) and Eagle (2021) provide nice brief introductions to these ideas; a more extensive one is Dasgupta (2011). 
 Standard textbooks are Calude (2002), Li and Vit\'a{}nyi (2008),  Nies (2009), and Downey and Hirschfeldt (2010). }
\begin{enumerate}
\item \emph{Patternless}:\footnote{
The main idea is that $x\in 2^\N$ is random--or, indeed, \emph{typical}--if it is \emph{not atypical}, where, as a first thought, $x$ would be atypical if it lies in some null set, i.e., $x\in\CN$ for some $\CN\in\Sg$ with $P(\CN)=0$. But this cannot work since $x\in\{x\}$ and $f^\N(\{x\})=0$ for any $x\in 2^\N$. Hence one should restrict the null sets $\CN$ relative to which $x$ is deemed atypical, and \ML's solution was to take all sets of the form $\CN=\bigcap_{N\in\N} U_N$, where the $U_N\subset X$ form 
 a countable number of suitably computable subsets  $U_N$ ($N\in \N$) for which $P(U_N)\leq 2^{-N}$ (where  we may assume that $U_{N+1}\subset U_N$), each of which states, informally, that $x$ contains some (computable) pattern; for example, if $X=2^\N$, one may think of $U_N=[0\cdots 0]$, with $N$ zeros, so that $x\in U_N$ iff $x$ starts with $N$ zeros.  Hertling and  Weihrauch (2003) and  Hoyrup and  Rojas (2009) extend this to arbitrary computable probability spaces.} 
 $x$ has \emph{no} pattern, where \emph{having} a pattern means  
$x\in \bigcap_{N\in\N} U_N$ for \emph{some} (\ML) \emph{test} $U=(U_N)_{N\in\N}$ (i.e.\ $U_N\subset 2^\N$,  $P(U_N)\leq  2^{-N}$,  and $U$ is suitably computable).\footnote{Different constructions of algorithmic randomness depend on the kind and the degree of computability one requires in various steps of the construction, for which we refer to Nies (2009) and Downey and Hirschfeldt (2010).}
 In other words,  $x$ is $P$-random iff for \emph{all} such tests $(U_N)$ one has  $x\notin\bigcap_{N\in\N} U_N$. 
\item \emph{Incompressible:} there is a constant $C\in\N$ such that for all $N$ we have 
\beq
K(x_{|N})\geq N-C, \label{ofm}
\eeq where $x_{|N}$ is the truncation of $x$ to its first $N$ digits, and $K(\sg)$ is  the Kolmogorov complexity of some \emph{finite} binary string $\sg$,
roughly speaking defined as the length of the shortest computer program that produces $\sg$, and more precisely 
defined by fixing some  universal prefix-free Turing machine $U$ (i.e., whose domain must be prefix-free),\footnote{ 
 Li and  Vit\'{a}nyi  (2008), chapter 3, explain why the theory does not work well for arbitrary Turing machines.}   and putting 
\beq
K(\sg)=\min_{\ta\in 2^*} \{\ell(\ta)\mid U(\ta)=\sg\}, \label{KC1}
\eeq
where $\ell(\ta)$ is the length (i.e., the number of bits) of $\ta$.
Thus  \er{ofm}, or more generally the property $K(\sg)\geq \ell(\sg)-C$,  indicates that the computer must already contain $\sg=x_{|N}$ in its memory in order to ``compute'' it;  or, in other words, that 
\emph{the  shortest computable description of  $\sg$ is $\sg$ itself.}  That is, $\sg$ is incompressible.
 \item \emph{Unpredictable:} no (left) c.e.\ (computably enumerable) betting strategy succeeds on $x$.\footnote{Here a  betting strategy is formalized as a martingale. For a good introduction to this aspect of algorithmic randomness see Nies (2009), Chapter 7.}
\end{enumerate}
Thus random initial conditions of the kind that have been proposed in (\emph{prima facie}) deterministic theories like Bohmian mechanics and kinetic theory,  which I have argued to provide a necessary sharpening of the typicality paradigm,  can at least be \emph{defined};\footnote{Defining what it means for $x\in X$ to be $P$-random should be distinguished from defining $x$ as (say) an infinite binary sequence or a real, see below.} but they are defined, in at least three equivalent ways,  by what they are \emph{not}. Hence the suspicion  arises that although one may be able to state what random initial conditions are \emph{not}, it may be difficult to say what they \emph{are}. 

And indeed, the elusiveness of randomness comes to a head with Chaitin's incompleteness theorems.\footnote{The original versions are spread out over a  number of papers and books by Chaitin, see e.g.\ Chaitin (1987, 2001),  surrounded by controversial  claims about the ubiquity of randomness in mathematics and the role of the complexity of axioms systems in this respect.
 Reliable sources for the theorems are van Lambalgen (1989), Raatikainen (1998, 2000), and Calude (2002), Theorems 8.4 and 8.7. One reason why  Chaitin's second incompleteness theorem has not   found its way into the philosophy of physics literature so far (as far as I know except  for Landsman, 2021) may be that Chaitin and others only stated it for the halting probabilities $\Om$, whereas it holds more generally, as discussed in the main text. \label{Omfn}} The second of these describes the situation in an idealized case like $X=2^\N$, which we prefer because of its greater clarity, whilst the first is concerned with finite binary strings. 
 In its original form, Chaitin's second theorem  states that in any sufficiently rich mathematical theory $T$ such as ZF(C),\footnote{As in G\"{o}del's incompleteness theorems, the theory $T$ is assumed to contain enough arithmetic (i.e., short of Peano arithmetic, at  least Robinson arithmetic); is recursively axiomatizable; and sound. To avoid trivial counterexamples to some of the claims below we assume that ZF(C) etc.\ has not been extended by axioms stating some of the things claimed to be unprovable. One may also take $T$ to include axioms for Bohmian mechanics or kinetic theory, etc.} the values of 
  all but finitely many digits  of his own random numbers $\Om$,  defined as the halting probabilities of certain (prefix) Turing machines,\footnote{These numbers coincide with the left computably enumerable (c.e.) random sequences, which form a very special subclass of the set of all random sequences in the sense of \ML. See e.g.\ Calude (2002), Chapter 7. \label{Omega} } are undecidable in the G\"{o}delian sense; that is, although all digits of $\Om\in 2^\N$ do have a value $x_n=v_n$ ``for God'', where $v_n\in \{0,1\}$, the set of all $n$ for which  $x_n=v_n$ can be \emph{proved} in $T$ is finite.
   However,   the first proof of the theorem in Calude (2002), Theorem 8.7, which I will summarize, obviously generalizes to any \emph{definable} \ML\ random sequence $x$ (so that $x_n=v_n$ is a well-defined sentence in first-order logic).\footnote{Here $x\in 2^\N$ is definable if, roughly speaking, it is given by a formula $x=\phv$
(interpreted in $2^\N$). This makes only a countable number of infinite binary sequences definable, among those the Chaitin $\Om$ numbers.}
  Namely,  defining randomness as patternlessness, a rough sketch 
  would be that if infinitely many  bits $x_n$ of $x\in 2^\N$ are known through some proof, we may define $U_N$ as the subset of all  $x$ in $2^\N$ that agree with the first $N$ of these. These $U_N$ form a \ML-test: the computability requirements on these tests are satisfied by the fact that (after G\"{o}del coding) the set of theorems is recursively enumerable, at least in ZF(C) etc. Since $x\in \bigcap_N U_N$ by construction, it cannot be random.

For non-definable (or indeed any) $x\in 2^\N$ a trick yields a similar result: add either a unary function symbol  $\ul{x}$ or a  countable number of  constants $\ul{x}_n$  ($n\in\N$) to $T$ with interpretation $x_n\in 2$ or $x\in 2^\N$ respectively, and once again, for any $n\in\N$ the expression $\ul{x}_n=v_n$ is a well-defined sentence in $T$. Ignoring this subtlety, I will continue to write this sentence as $x_n=v_n$. We then have:
\begin{theorem}
 In any sufficiently rich axiomatic  and sound mathematical framework $T$: if $x\in 2^\N$ is (\ML) random, then 
 only finitely many sentences $x_n=v_n$ are decidable in $T$.
\end{theorem}
The proof is the same as for definable sequences $x$, \emph{mutatis mutandis}, and duly generalizes to arbitrary $P$-random points in computable probability spaces $(X,\Sg,P)$ whose points correspond to reals (so that one can talk about digits of binary or decimal expansions) and whose probability measure is equivalent to Lebesgue with a bounded density, to be on the safe side.

Returning to $X=2^\N$ with $P=f^\N$ (i.e., to \ML\ randomness in its original form),  defining randomness as incompressibility gives a rich picture of incompleteness that is also useful for my argument because it shows how the idealization to infinite (binary) sequences in $2^\N$ relates to randomness of finite (binary) strings in $2^*$. 
  \emph{Chaitin's first incompleteness theorem} states that: 
  \begin{theorem}\label{t32}
For any sufficiently rich and sound theory $T$, there is a constant $c\in \N$ (depending only on $T$ and the choice of the machine used to define $K$) such that $T$
  can  prove no statement of the form $K(\sg)\geq  c$ (although the statement is true for all but finitely many binary strings $\sg\in 2^*$).\end{theorem}
  We in fact do not  need this theorem itself, but only a corollary of it. For fixed $C\in\N$, say that $\sg\in 2^*$ is \emph{$C$-random} if $K(\sg)\geq \ell(\sg)-C$, and \emph{incompressible} if it is $0$-random, that is, $K(\sg)\geq\ell(\sg)$.
  \begin{corollary}\label{C1} 
  \begin{enumerate}
\item   For fixed $C\in\N$  one can prove $C$-randomness of $\sg$ 
 for only finitely many $\sg\in 2^*$  (although for any $N\geq C$  at least $2^N(1-2^{-C})$ elements $\sg\in2^N$ are $C$-random).
 \item If $\sg\in 2^*$ is incompressible, then this cannot be proved beyond some finite value of $\ell(\sg)$.
 \item  If $x\in 2^\N$ is random, then  the sentence \er{ofm} can be proved, from given initial segments $x_{|N}$, only for finitely many values of $N$ (although the sentence is true for all $N$).
\end{enumerate}
\end{corollary}
The  sentence between brackets in part 1 of this corollary follows from a simple counting argument,\footnote{See  Gr\"{u}nwald and Vit\'{a}nyi (2008),  \S2.2.} which also implies the  sentence between brackets in Theorem \ref{t32}.
Here is a proof sketch of part 1 (which trivially  implies parts 2 and 3).\footnote{See Gr\"{u}nwald and Vit\'{a}nyi (2008), \S4.3, Example 4, slightly adjusted using Raatikainen (2000).}
  Since  theorems are recursively enumerable, 
 there exists a program $\pi$ (or computable function), running on 
the  machine $U$ used to define the Kolmogorov complexity $K(\cdot)$,  that for any input $(N,C)$ halts iff it hits the first theorem of the form
\beq
K(\sg)\geq N-C, \label{KNC}
\eeq
 for some $\sg\in 2^N$ and $C\in\N$,  and then outputs $\pi(N,C)=\sg$. As we have seen, such strings $\sg$ \emph{exist}; what is at stake here is if they can be found through \emph{proof}. 
 By definition of  $K(\cdot)$, this gives 
\beq
K(\sg)\leq \ell(\pi)+O(\log_2(N)) +O(1),\label{ellell}
\eeq
 where $\ell(\pi)$ is the length of the program $\pi$ (in binary code) and $\log_2(N)$ 
 is what $\pi$ needs to print $\sg\in 2^N$.
If this could be done for arbitrary $N$ one would, for sufficiently large $N$, obtain a contradiction between 
 \er{ellell}, which follows if \er{KNC}  were provable, and the latter inequality itself.\QED
     \section{Discussion and conclusion}\label{S4}
We return to our initial conditions $(z_0, z_1, \ldots)$, where $z_n\in\R^3$ or $z_n\in \R^6$ in our case studies.
But it is enough for my philosophical purposes to take $z_n\in 2$; it is the randomness of the configuration of the particles \emph{relative to each other} that counts, rather than the additional randomness inherent in using real variables for each particle.\footnote{Doing so would add even more unpredictability and randomness, cf.\ Del Santo and Gisin (2025).
In algorithmic Bohmian mechanics \`{a} la Barrett, Chen, and Lopez-Wild (2026ab) one keeps $N$ finite, but takes the measurement sequence to be infinite to get into the \ML\ framework. I replace their $X=\R^{3N}$ by $X=2^\N$ and, following Landsman (2023), take the infinite sequence of measurements to simply consist of reading the initial condition $x\in 2^\N$.} Thus my initial conditions are simply
 taken to be  infinite binary sequences $x\in 2^\N$ (equipped with probability measure  $P=f^\N$, the unbiased Bernoulli measure,  as usual), the underlying single degrees of freedom being bits rather than particles.
 
  By the previous section,  if $x$ is a ``God-given'' random initial condition (even, in Bohmian mechanics, of the universe), then $T$ can 
  accommodate only finitely many of its digits (for otherwise $x$ would not be random). 
To estimate this number,\footnote{Hoyrup (2020), \S2.2, gives a useful brief survey of randomness deficiencies, amplified by  e-mail correspondence.  These can be defined for arbitrary computable probability spaces using \ML's definition of randomness. That requires a universal \ML\ test $U$, which (re)defines $d_P(x)$ as the minimal value of $N\in\N$ such that $x\notin U_N$.} for any $x\in 2^\N$,  let the \emph{randomness deficiency} $d_P(x)$ be the minimal integer $C$ for which \er{ofm} holds for all $N$; then, by the incompressibility criterion in \S\ref{S3}, $x$ is random iff $d_P(x)<\infty$, in which case 
  the smaller $d_P(x)$ is, the ``more random'' is $x$. For a random sequence $x$, the number of bits $T$ can decide without contradicting the randomness of $x$ is  at most $N=d_P(x)+ O(1)$, where the $x$-independent $O(1)$ term depends on the mathematical framework $T$ (here taken to include the physical theory in question). For all other bits, God can, but $T$ cannot tell their value.  This supposes that $x$ is given in advance (by God or Nature). 
  
  Alternatively, the initial condition $x$ may be approximately \emph{chosen} by stating finitely many digits $x_0x_1\cdots x_{N-1}$. If this process continues in some systematic way (i.e., is effective),\footnote{If the process is random  it leads with probability one  to a random sequence, but it cannot be proved that it does, and one can hardly maintain that initial conditions are \emph{chosen} in such a way.
} then
   the larger $N$ becomes, the larger the randomness deficiency $d_P(x)$  of any possible limit $x$ grows with $N$, and hence, by the previous argument, the less random the limit will be; until, as $N\raw\infty$, the limit sequence cannot be random at all. In other words, by trying to be increasingly precise, our well-intending physicist in fact 
  increasingly damages the goal of stating a random initial condition, pushing $x$ towards the edge of the subset of random sequences and thus severely limiting the allowed class of random initial conditions $x$ to those with large randomness deficiency.\footnote{One could  try different approximation procedures than just increasing the number of digits, but any effective procedure to approximate a random sequence $x$ with a computable rate of convergence must fail,  for otherwise $x$ could not be random. In other words, the rate of convergence of  approximations of $x$ by computable numbers is necessarily unknown.
 In at most a countable number of cases, $x$ may be specified by some logical formula (that is, $x$ is definable), such as Chaitin's $\Om$ numbers (see footnote \ref{Omega}),  but even in the best cases a computable (left) approximation to $x$ with necessarily noncomputable 
rate of convergence converges to $x$ in the slowest possible way (cf.\ Calude, 2002, Scholium 7.63 and surrounding text), and most definable $x$ have no computable approximation at all. If the initial condition is not even definable, which is the ``typical'' case (!), it is even unclear what it is that one approximates.} 

Here, my infinite idealization, which enables Chaitin's \emph{second} incompleteness theorem, is used to state sharp results (as in e.g.\ theories of phase transitions and/or symmetry breaking), but it 
 is not necessary; in finite systems, say with $N<\infty$ particles,  we could use Chaitin's \emph{first} incompleteness theorem.
Replacing randomness of infinite sequences $x$ by incompressibility of sufficiently long finite strings $\sg$, the finite shadow of  the first point above is that if  $\sg$ has been proved to be incompressible, then its digits cannot be explicitly  stated; the second point corresponds to trivial fact that if the digits of $\sg$ are stated, then $\sg$ will not be incompressible.

These problems with random initial conditions obviously have consequences for the ensuing trajectories. The Bohmians evidently believe that these are unproblematic, for example:    \begin{quote}
\begin{small}
 Now we come to the trajectories. We have a theory of particles in motion and we are
now eager to compute them in situations of interest. Just as we compute the orbits
of planets, we would like to compute the orbits of the Bohmian particles around
the nucleus in an atom. Why should we be eager to do that? Because the theory
allows us to do that? So the electron moves around the nucleus in some way. Is
that exciting? Yes, you may say, because all kinds of exciting questions come up. (D\"{u}rr and Teufel, 2009, p.\ 151)
\end{small}
\end{quote}
The analogy with planetary orbits is instructive.
One could argue that any deterministic theory with continuous degrees of freedom faces the  problem that almost every point is random (with respect to some normalized version of Lebesgue measure). But it seems special to kinetic theory and Bohmian mechanics that one \emph{must} use random initial conditions.\footnote{According to Barrett, Chen, and Lopez-Wild (2026ab) this randomness should even be seen as a Law of Nature. }
 Other dynamical theories can use arbitrary computable (even rational) points in $\R^n$ to approximate any desired initial condition without any problem. Kinetic theory and Bohmian mechanics do not have this  flexibility. 
  
To make the point  differently: mathematical physics provides advanced mathematical theories to explain, interpret, understand, and manipulate what van Fraassen (2008) calls
\emph{data models}, which collect observations in low-grade mathematical form like numerical tables. The incompleteness argument shows that  Bohmian mechanics and kinetic theory with random microscopic initial conditions cannot process such data, except by averaging them out. For kinetic theory this is hardly a problem since its goal is to describe macroscopic properties and the rest is conversation,
but for Bohmian mechanics  it moves the layer it adds to quantum mechanics beyond physics.

I see three potential ways out, each  unattractive in my view, but here they are:
\begin{enumerate}
\item   
One trusts that any finite information about \emph{some} initial condition (given or stated)  is information about a \emph{random} initial condition. This is typically true: with high probability for finite $N$, and with unit probability for $N=\infty$. Thus one relies on the old typicality reasoning, and therefore in this scenario the point of introducing $P$-randomness seems to be lost.

 But this  also seems unacceptable:
\item If one returns to the original typicality reasoning behind Bohmian mechanics and kinetic theory, supported by probability theory so as to obtain a precise notion of typicality but not reinforced by algorithmic randomness, then, as argued throughout this paper:\footnote{See also Barrett, Chen, and Lopez-Wild (2026ab) for convincing arguments in this direction.}
\begin{enumerate}
\item it remains unclear what it means to be ``randomly distributed'';
\item the theory says nothing about individual systems, or about \emph{this} world. 
\end{enumerate}
The theory only talks about `typical' worlds. This is much, but not all we could go for. 
\item Accepting the need for random initial conditions including what this means mathematically, one may accept that the particle trajectories are really there in the world, but happen to be inaccessible not only empirically but even to Bohmian mechanics itself. This may sound similar to the idea,  in G\"{o}del's setting, that undecidable number-theoretic statements are actually true (or false) about natural numbers ``in reality'', but  merely happen to be inaccessible to some  given axiomatic framework in which they are undecidable (like ZFC or PA).  
Likewise, 
 Bohmian mechanics may be claimed to be an accurate description of ``reality''.
 
I would object to this argument that the said interpretation of G\"{o}del's incompleteness theorems  hinges on a highly nontrivial and controversial assumption about mathematics, namely the belief that any proposition about at least natural numbers is either true or false \emph{full stop}, i.e., irrespective of axioms, deductions, etc.\footnote{This 
is called \emph{truth-completeness of arithmetic} (Paseau and Pregel, 2023), who also call it  a `\emph{fundamental commitment of mathematics}'. It has been well expressed by Hardy (1929, p.\ 4): `\emph{Any number is the sum of 4 squares; any number is the sum of 3 squares; any even number is the sum of 2 primes}. These are not convenient working hypotheses, or half-truths about the Absolute, or collections of marks on paper, or classes of noises summarising reactions of laryngeal glands. They are, in one sense or another, however elusive and sophisticated that sense may be, theorems concerning reality, of which the first is true, the second is false, and the third is either true or false, though which we do not know.'
See also Landsman \& Singh (2025) for a critique of this attitude (which is shared by most mathematicians).} In my view this idea has been made very unattractive by the critique of Dummett (1963/1978).
Likewise, this defense of Bohmian mechanics presupposes some kind of realism, but it would be of a very strange kind compared even with the above natural number realism, since whatever the precise relationship between numbers and  number \emph{theory}, numbers themselves are all around us, just like particle trajectories. But the \emph{Bohmian} trajectories are nowhere to be seen as a matter of principle.

  Even if one accepts this kind of realism, the analogy with G\"{o}del breaks down, because undecidable statements in number theory are rare in practice and almost all of mathematics can be done without them.\footnote{Chaitin's denial that such statements are rare is controversial, see 
  Calude and J\"{u}rgensen (2005) for a precise result.}
   In contrast, Bohmian mechanics is \emph{all about} the trajectories--their alleged existence marks the decisive difference with conventional \qm. In other words, G\"{o}del's theorems do not undermine our ability to discuss almost all theorems about numbers; but the undecidability of random initial conditions in Bohmian mechanics incapacitates our ability to talk about not just some, but  \emph{all} 
 of its particle trajectories.
\end{enumerate}

The case of kinetic theory is slightly different: here, standard lore has it that in classical statistical mechanics one has to resort to probability theory because initial conditions and the ensuing particle trajectories are known in principle but not in practice.  Now once again, as in Bohmian mechanics, on the necessary randomness assumption pinning down these initial conditions, the trajectories are not even known in principle, at least not to the theory supposed to describe them. The difference is that in kinetic theory no one doubts that there are such trajectories, with 
 specific numerical values; but once again these values are beyond the reach of the theory in question.

Let me finally muse on the implications of this analysis for determinism.
 As a  simplification that causes no harm I use discrete time, so that our  dynamics is given by a  (measurable) map 
 \beq 
 f: 2^\N\raw 2^\N;
 \eeq
  the state of the system at time $t\in\N$ is then given by $f^t(x)$. 
  Moreover, I assume the arguably most favourable scenario for  Bohmians and other devoted determinists, namely that  $f$ is computable.\footnote{The following discussion applies equally well  to the weaker (hence less favourable) assumption that $f$ is layerwise computable, as proposed by Barrett and Chen (2025), since such functions are computable at random points. For the trajectories to be undecidable as a consequence of their initial conditions being undecidable some assumptions are necessary to avoid trivial cases like the dynamics being a constant map.
 For example, it is sufficient that the dynamics preserves randomness, as is  the case in Bohmian mechanics for layerwise computable maps (Barrett, Chen, and Lopez-Wild, 2026a) as well as for the maps used in toy models of kinetic theory (Dekkers and Landsman, 2026).
 } 
 In that case the computation of $f(x)$ for any $x\in 2^\N$ proceeds bit by bit,\footnote{See Weihrauch (2000), Chapter 2. In particular, by Lemma 2.1.11 computable functions $f:2^\N\raw 2^\N$ correspond to monotone computable functions $\til{f}: 2^*\raw 2^*$ (i.e., $\sg\prec\ta$ implies $\til{f}(\sg)\prec\til{f}(\ta)$) in that $f(x)$ is the unique sequence such that $\til{f}(x_{|N})\prec f(x)$ for all $N\in\N$. Here $\sg\prec\ta$ means that $\sg$ is a prefix of $\ta$, that is, $\ta=\sg\cdots$. This simply means that a computable function $f$ first computes $\til{f}(x_0)$, then $\til{f}(x_0x_1)$, which adds a finite string to $\til{f}(x_0)$, etc. 
 } but if $x$ is a  (\ML) random sequence, then the bitstream cannot be computed from anything $T$ can supply, and hence the trajectories cannot be computed. What remains of determinism in that case?

First, I agree with Butterfield (2005), \S 2, that `The key idea is that determinism is a property of a theory.'\footnote{My motivation for agreeing with this is my belief that  theories of mathematical physics are not \emph{descriptions of} the world but Wittgensteinian yardsticks or objects of comparison \emph{held against} the world (Landsman, 2026), but one need not align with that in order to agree with Butterfield (and mainstream philosophy of physics). Either way I avoid Earman's  possible world semantics in concepts of determinism.
See also  M\"{u}ller and Placek  (2018) and  Hoefer (2026).} Let me initially also follow his definition of determinism, according to which:
\begin{definition}\label{JBD}
A theory is deterministic if and only if: any two of its models that agree at a time $t$ on the state of their objects, also agree at all times future to $t$, where a \emph{model} of the theory is defined as a sequence of states  that conforms to the laws of the theory.\footnote{If the laws are partial differential equations, then well-posedness including uniqueness of solutions often relies on specific regularity properties of the state space $X$. For the simple model $X=2^\N$  this doesn't matter, though.}
\end{definition}
These laws  include the dynamical map $f:X\raw X$, where $X$ is the state space of the theory. But  in Bohmian mechanics, and arguably  in kinetic theory, too, the laws, exceptionally,  also include $P$-randomness of the initial state  with respect to some  probability measure $P$ on the state space $X$.
But, as we have just seen, since in that case the initial conditions cannot be specified, 
 the numerical values of these trajectories cannot be specified either, not even to finite but arbitrary precision.  They were already hidden in physics; now they are even hidden in mathematics.\footnote{
Pictures of trajectories one often sees in Bohmian mechanics literature must have been computed from finite data whose connection to truly random initial data is obscure, as explained at the beginning of this section.}
In my view, this admittedly  special situation calls for a modification of Butterfield's (2005)  definition of determinism just quoted. To deal with possible undecidability  I propose the following addition: 
\begin{definition}\label{KLD}
A theory $T$ with state space $X$ is deterministic iff it is deterministic in the sense of Definition \ref{JBD}  \emph{and}
 $T$ is able, at least in principle,  to approximate lawful sequences of states with arbitrary precision from lawful initial conditions stated with a precision enabling this. 
\end{definition}
 The idea of this addition is that $T$ is now also in principle able to decide the outcomes of  experiments, since this relies on its ability to process numerical (initial) data; in fact, my addition could be replaced by this ability (but in that case one would need to define ``experiments'').  Then:
\begin{itemize}
\item  Bohmian mechanics satisfies Definition \ref{JBD} but fails Definition \ref{KLD}, at least if the initial condition $x$ is lawful and  God-given, so that it has a given deficiency index $d_P(x)$.\footnote{This is surely how the theory is intended, but my argument also works if $x$ is effectively stated by a physicist, since then either one has limited accuracy or $x$ cannot be random. This adds to a previous argument for the indeterminism of Bohmian mechanics, to the effect that its need to invoke an oracle given by a random sequence seems incompatible with its determinism (Landsman, 2022). See also  Oldofredi (2023) and van Strien (2024) for interesting information about Bohm's own views on determinism in general--the title of the last paper is self-explanatory!} 
\item So does ``$P$-randomized'' kinetic theory.\footnote{This still maintains its practical power, since the description of microscopic  trajectories was never its purpose.
}
\item  Newtonian mechanics and the \textsc{pde}s of classical physics (at least with a well-posed initial value problem) satisfy both,\footnote{The example by Pour-El and Richards (1989), pp.\ 68--69,  of a wave equation with computable initial data but non-computable solution does not upset this, since the solution is known analytically and can be given exactly. This case is in fact quite opposite to random initial conditions, which are as far from computable as possible.}
 as does quantum mechanics without measurement; in all these cases one can freely increase one's digital precision in either reading or providing initial data, since these are not bound by any law (like being random) preventing this. 
\item  Copenhagen \qm\ fails both definitions;
   the role of G\"{o}delian undecidability in Bohmian mechanics is played by a more informal kind of  undecidability in describing outcomes of measurements (and even of the measurement process itself).\footnote{Bohmian mechanics moves this undecidability back from the  outcomes to the initial conditions it has added to the ontology.
The results of Chen and Tumulka (2026) to the effect that apart from the particle trajectories even the wave function of the universe is fundamentally unknowable confirm my impression that what remains of Bohmian mechanics is some kind of a black box whose alleged ontological content  is inaccessible, unknown, unknowable, and undecidable.} 
\end{itemize}

Finally, in view of all this it might  be worth looking for a definition of $P$-randomness that does not stumble upon some incompleteness theorem, to which end the whole algorithmic approach to randomness should probably be abandoned.\footnote{Barrett and Chen (2025) consider the case where \ML\ randomness is replaced by what is called Schnorr randomness, concluding that these cannot be distinguished empirically. Moreover, both Schnorr randomness and the even weaker concept of Kurtz randomness fall victim to Chaitin's theorem, and randomness concepts even weaker than Kurtz's are unacceptable for various reasons (e.g.\ allowing c.e.\ sequences).
Something beyond the scope of algorithmic randomness seems necessary, in particular,  a theory not based on the theory of computation (i.e.\ recursion theory).} But it is questionable whether this could work even in principle, 
 since its elusiveness will always be a fundamental property of randomness.

\addcontentsline{toc}{section}{References}
\begin{small}

\end{small}
\end{document}

%% file: Bohrificationpreamble.tex
\newcommand{\id}[1]{\ensuremath{\mathrm{id}}}

\newcommand{\half}{\mbox{\footnotesize $\frac{1}{2}$}}

\newcommand{\qm}{quantum mechanics}
\newcommand{\er}{\eqref}
 
\newcommand{\beq}{\begin{equation}}
\newcommand{\eeq}{\end{equation}} 
\newcommand{\bea}{\begin{eqnarray}}
\newcommand{\eea}{\end{eqnarray}}

\newcommand{\ul}{\underline}

 \newcommand{\til}{\tilde}
\newcommand{\raw}{\rightarrow}

\newcommand{\lraw}{\leftrightarrow}

\newcommand{\dl}{\delta} 
 \newcommand{\varep}{\varepsilon}

\newcommand{\lm}{\lambda} 
 \newcommand{\sg}{\sigma}
\newcommand{\Sg}{\Sigma} \newcommand{\ta}{\tau} 
 \newcommand{\phv}{\varphi}
\newcommand{\ch}{\ch}  
 \newcommand{\Om}{\Omega}

\newcommand{\CN}{{\mathcal N}}

\newcommand{\N}{{\mathbb N}} \newcommand{\R}{{\mathbb R}}
 
  \makeatletter
 
\makeatletter
\def\moverlay{\mathpalette\mov@rlay}
\def\mov@rlay#1#2{\leavevmode\vtop{%
   \baselineskip\z@skip \lineskiplimit-\maxdimen
   \ialign{\hfil$\m@th#1##$\hfil\cr#2\crcr}}}
\newcommand{\charfusion}[3][\mathord]{
    #1{\ifx#1\mathop\vphantom{#2}\fi
        \mathpalette\mov@rlay{#2\cr#3}
      }
    \ifx#1\mathop\expandafter\displaylimits\fi}
\makeatother